\documentstyle[eqsecnum,aps,epsf]{revtex}

\newcommand{\eq}[1]{(\ref{#1})}

\newcommand{\beq}{\begin{equation}}
\newcommand{\eeq}{\end{equation}}
\newcommand{\beqn}{\begin{eqnarray}}
\newcommand{\eeqn}{\end{eqnarray}}

\def\NP{Nucl.~Phys.}
\def\PL{Phys.~Lett.}

\def\PR{Phys.~Rev.}
\begin{document}
\draft
\preprint{KANAZAWA 97-10}
\title{Abelian Monopoles in $SU(2)$ Lattice Gauge Theory\\
as Physical Objects}
\author{B.L.G.~Bakker}
\address{Department of Physics and Astronomy, Vrije Universiteit,\\
De Boelelaan 1081, NL-1081 HV Amsterdam, The Netherlands}
\author{M.N.~Chernodub and M.I.~Polikarpov\cite{Kanazawa}}
\address{ITEP, B.Cheremushkinskaya 25, Moscow, 117259, Russia}
\date{\today}
\maketitle
\begin{abstract}
By numerical calculations we show that the abelian monopole currents
are locally correlated with the density of the $SU(2)$ lattice
action. This fact is established  for the maximal abelian projection.
Thus, in the maximal abelian projection, the monopoles are  physical
objects, they carry the $SU(2)$ action. Calculations on the
asymmetric lattice show that the correlation between monopole
currents and the density of the $SU(2)$ lattice action also exists in
the deconfinement phase of gluodynamics.
\end{abstract}

\widetext

\section{Introduction}

The monopoles in the maximal abelian projection (MaA projection) 
of $SU(2)$ lattice gluodynamics \cite{KrScWi87} seem to be 
responsible for the formation of the flux tube between the test 
quark-antiquark pair. The $SU(2)$ string tension is well described  
by the contribution of the abelian monopole currents 
\cite{ShSu94,StNeWe94,BaBoMu96}  which satisfy the
London equation for a superconductor
\cite{SiBrHa93}. The study of monopole creation operators shows
that the abelian monopoles are condensed \cite{DiGi95,Nak96,ChVePo96}
in the confinement phase of gluodynamics.

On the other hand,  the abelian monopoles arise in the
continuum theory \cite{tHo81} from the singular gauge transformation
and it is not clear whether these monopoles are ``real'' objects. A
physical object is something which carries action and in the present
publication we only  study the question if there are any correlations
between abelian monopole currents and $SU(2)$ action. In
 \cite{ShSu95} it was found that the total action of $SU(2)$
fields is correlated with the total length of the monopole currents,
so there exists a global correlation. Below we discuss the local
correlations between the action density and the monopole currents.

\section{Correlators of Monopole Currents and Density of $SU(2)$ Action}

The simplest quantity which reflects the correlation of the local
action density and the monopole current is the relative excess of
$SU(2)$ action density in the region near the monopole current. It
can be defined as follows. Consider the average action  $S_m$  on the
plaquettes  closest to the monopole current $j_\mu(x)$. Then
the relative excess of the action is

\beq
\eta = \frac{S_m - S}{S}\,, \label{eta}
\eeq
where $S$ is the standard expectation value of the lattice action, $S
= <\left( 1 - \frac 12 Tr \, U_{P}\right)>$. $S_m$ is defined as
follows:
\beq
S_m= <\frac 16
\sum_{P \in \partial C_\nu (x)}
\left( 1 - \frac 12 Tr \, U_P \right)>\, , \label{Smlat}
\eeq
where the average is implied over all cubes $C_\nu (x)$
dual to the magnetic monopole currents $j_\nu(x)$, the summation is
over the plaquettes $P$ which are the faces of the cube $C_\nu (x)$;
$U_P$ is the plaquette matrix. For the static monopole  we have $j_0 (x) \neq
0,\, j_i (x) =0, \, i=1,2,3$,  and only magnetic part of $SU(2)$
action density contributes into $S_m$. The correlation of the
monopole currents and the electric part of the action (which comes from
more distant plaquettes)  will be  studied in another  publication.

At large values of $\beta$,  the quantity $\eta$ is equal to the
normalized correlator of the dual action density and the monopole current:
\beq
C = \frac{<\frac 12 Tr\,\left( j_\mu (x)\,
\tilde{F}_{\mu\nu}(x)\right)^2 >}
{<j_\mu^2 (x)> <\frac 12 Tr F_{\alpha\beta}^2(x)>} - 1 \,. \label{C}
\eeq
Here  the lattice
regularization is  implied, in particular,
\begin {eqnarray*}
&\left<\frac 12 Tr F_{\alpha\beta}^2(x)\right>=
\left<\left( 1 - \frac 12 Tr \, U_{P}\right)\right>\,,&\\
&\left<\frac 12 Tr\,\left( j_\mu
(x)\, \tilde{F}_{\mu\nu}(x)\right)^2\right>=
\left<\sum\limits^4_{\mu=1} j_\mu^2(x) \cdot \frac 16
\sum\limits_{P \in \partial C_\mu (x)} \left( 1 - \frac 12 Tr \, U_P
\right)\right>\,,
\end {eqnarray*}
 the notations are the same as in \eq{Smlat}. In the MaA
projection at sufficiently large values of $\beta$,  the probability of $j_\mu
(x) = \pm 2$ is small. From the definitions \eq{eta}, \eq{Smlat} and \eq{C},
it follows that if $j_\mu (x) = 0, \pm 1$, then  $\eta = C$. Numerical
calculations show  that $\eta = C$ with the  accuracy of 5\% for $\beta >
1.5$ on lattices of sizes $10^4$ and $12^3 \cdot 4$.

\section{Numerical Results}

We calculate the quantities $\eta$ and $C$ on the symmetric $10^4$ lattice
and on $12^3 \cdot 4$ lattice which corresponds to finite
temperature.  In both cases, it occurs that in the MaA projection we have
 $\eta \neq 0$ and $C \neq
0$ for all values of $\beta$.  We also
consider the abelian projection  which corresponds to the
diagonalization of the plaquette matrices in the 12 plane (the $F_{12}$
gauge) and  the diagonalization of the Polyakov line (the Polyakov
gauge).

\begin{figure}[htb]
\centerline{\epsfxsize=0.95\textwidth\epsfbox{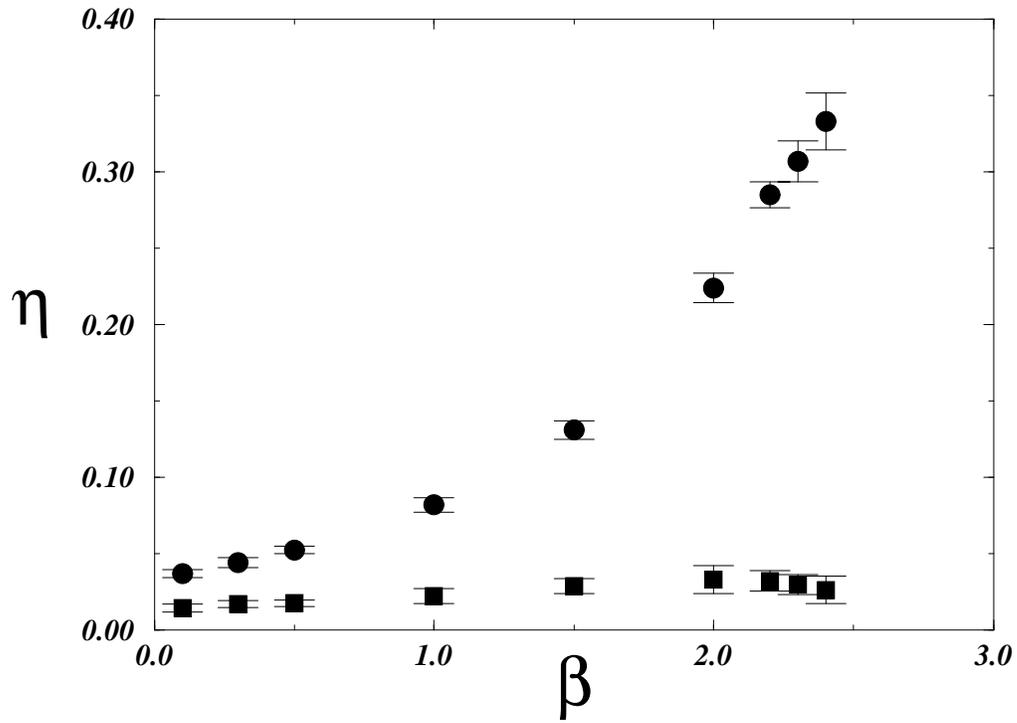}}
\vspace{-5cm}
\caption{The relative excess of the magnetic action density near the
monopole current, $\eta$, for the $10^4$ lattice. Circles correspond
to MaA projection, squares correspond to Polyakov gauge.}
\label{fig1}
\end{figure}

In Fig.~\ref{fig1} we show  the dependence of the quantity $\eta$ on
$\beta$ for $10^4$ lattice for the MaA projection and for the
Polyakov gauge. It turns  out that the data for the $F_{12}$
projection coincide within  statistical errors with the data for the
Polyakov gauge and we do not show  these.  In Fig.~\ref{fig3} we plot
the same data, this time, for  the $12^3\cdot 4$ lattice. It is seen
that the quantity $\eta$ is much smaller for the Polyakov gauge than
that for the MaA projection; the deconfinement phase transition at
$\beta \approx 2.3$ does not have much  influence on the behavior of
$\eta$. Thus, the monopole currents in the MaA projection are
surrounded by plaquettes which carry the values of $SU(2)$ action
larger than the value of the average action.

To obtain these results we consider 24 statistically independent
configurations of $SU(2)$ gauge fields for $\beta \le 2.0$, 48
configurations for $2.25 \le \beta \le 2.35$, and 120 configurations
for $\beta \ge 2.4$. To fix the MaA projection we have  used the
overrelaxation algorithm \cite{MaOg90}. The number of the gauge fixing
iterations is determined by the criterion given in
~\cite{Pou97}: the
iterations are stopped when the matrix of the gauge transformation
$\Omega (x)$  becomes close to the unit matrix: ${\rm max}_x \{ 1- {\frac 12}
Tr \,\Omega (x) \} \le 10^{-5}$.
It has been  checked that more accurate gauge fixing does not change our
results.

The correlation of the currents and the action density can be explicitly
visualized. In Fig.~\ref{fig3} we show  the  ``time'' slice of
$10^4$ lattice. The monopole currents are represented by lines (or by
large dots, if the current is perpendicular to the time slice). The
monopole currents are obtained in the MaA projection from the gauge
field configurations generated at $\beta = 2.4$. The density of the
small dots is proportional to $S(x)\,\theta\left( S(x) - S_c\right)$;
the action density is defined as usual: $S(x) = \sum_{\mu\nu}\left( 1
- \frac 12 Tr \, U_{\mu\nu}(x)\right)$. In Fig.~\ref{fig3} we have
$S_c = 0.75 <S(x)>$. For this value of the threshold $S_c$,  the
correlation has been found to be   most conspicuous\footnote{The
fluctuations of $S(x)$ are of the order $0.3<S(x)>$. For  the
threshold $S_c< (0.5<S(x)>)$,  the small dots superimpose on each
other in Fig.~\ref{fig3}; for $S_c > (0.85<S(x)>)$,  the density of
the small dots is small and the correlations are unclear. Actually,
Fig.~\ref{fig3} is just an illustration;  the existence of the
correlations of the currents and the action density is obvious since
$\eta > 0$, see Figs. 1,2.}.  In Fig.~\ref{fig3}  one can  see  some
currents which are not surrounded by small dots.  This indicates
that near these currents we have  $S(x) \le S_c$. Moreover,  there
are some regions with  high density of the action which are not
related to the monopole currents.  Inspecting several gauge field
configurations, we have found that in most cases these regions are
related to closed monopole currents  in the neighboring time slice.
At $\beta=2.4$, approximately 30\% of the regions with  high action
density are not explicitly related to  the monopole currents.

Thus we have found   that,  in the MaA projection,
 the {\it abelian} monopole currents and the regions with an excess of
the  {\it nonabelian} action density are
spatially correlated. We conclude that the monopoles in the MaA projection
carry action and thus constitute physical objects. It does not mean that
these  have to propagate in the Minkovsky space;  a  chain of instantons can
produce a  similar effect:  an  enhancement of the action density along a
line in  Euclidean space. It is important to understand what is the general
class of configurations of $SU(2)$ fields which generate  monopole
currents. Some specific examples are known, in particular, the  instantons
\cite{GuCh95,HaTe96,SUSTM97,BrOrTa97,FeMaTh97} and the  BPS--monopoles (periodic
instantons) \cite{SmSi91}. This question can be reformulated in another way:
are there any continuum physical objects which correspond to abelian
monopoles obtained in the MaA projection?

\acknowledgments

The authors are grateful to Prof. T.~Suzuki and Prof. Yu.A.~Simonov
for useful discussions. M.I.P. and M.N.Ch. feel much obliged for the
kind reception given to them by the staff of the Department of
Physics of the Kanazawa University  and by the members of the
Department of Physics and Astronomy of the Free University at
Amsterdam. This work was supported by the JSPS Program on Japan --
FSU scientists collaboration, by the grants INTAS-94-0840,
INTAS-94-2851, INTAS-RFBR-95-0681 and RFBR-96-02-17230a.

\begin{figure}[htb]
\centerline{\epsfxsize=0.95\textwidth\epsfbox{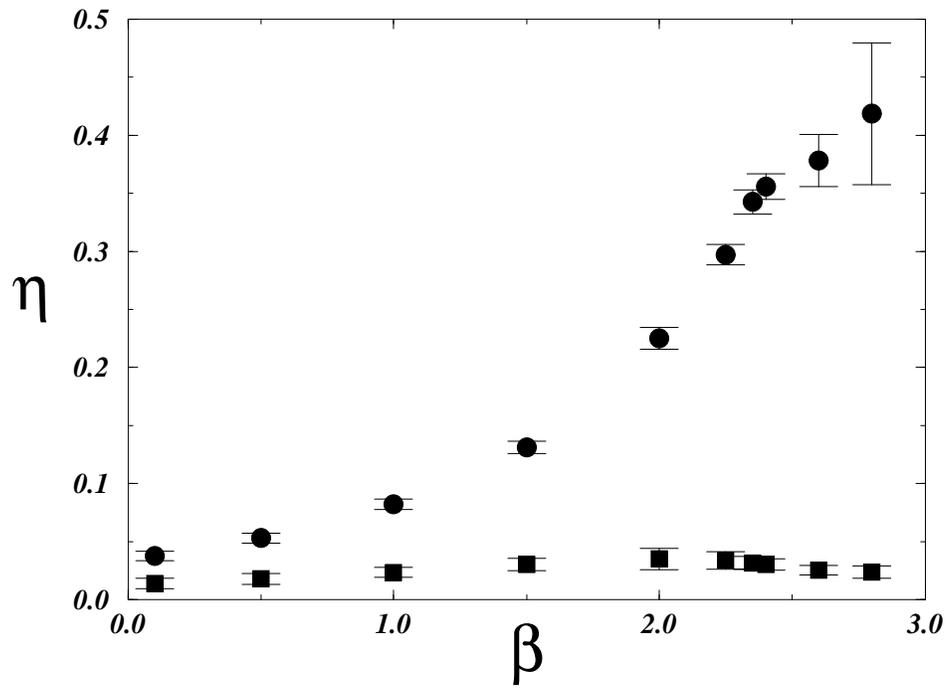}}
\vspace{-5cm}
\caption{The same as in Fig.~\ref{fig1}, but for the asymmetric
lattice $12^3 \cdot 4$.}
\label{fig2}
\end{figure}

\begin{figure}[htb]
\centerline{\epsfxsize=1\textwidth\epsfbox{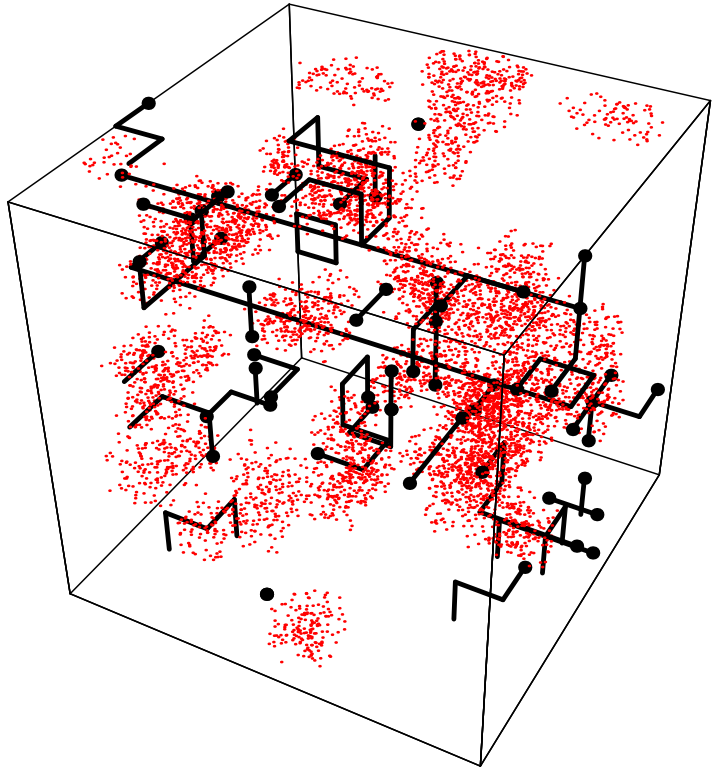}}
\vspace{-2cm}
\caption{Three dimensional slice of the four dimensional $10^4$
lattice. The lines and the big dots mark the monopole currents, the
density of the small dots is proportional to $SU(2)$ action
density.}
\label{fig3}
\end{figure}

\end{document}